\documentclass{pasa}%

\title[Spectral resolving power]{Quantifying resolving power in astronomical spectra} 
\author[J.G. Robertson]{J. Gordon Robertson$^{1,2}$\thanks{G.Robertson@physics.usyd.edu.au}\\  
\affil{$^1$Sydney Institute for Astronomy, School of Physics, University of Sydney, NSW 2006, Australia}%
\affil{$^2$Australian Astronomical Observatory, PO Box 915, North Ryde, NSW 1670, Australia}}%
\jid{PASA}
\doi{10.1017/pas.\the\year.xxx}
\jyear{\the\year}

\begin{document}%
\begin{abstract}
The spectral resolving power $R=\lambda / \delta \lambda$ is a key property
of any spectrograph, but its definition is vague because the `smallest
resolvable wavelength difference' $\delta \lambda$ does not have a
consistent definition. Often the FWHM is used, but this is not consistent
when comparing the resolution of instruments with different forms of
spectral line spread function.  Here two methods for calculating resolving
power on a consistent scale are given. The first is based on the principle
that two spectral lines are just resolved when the mutual disturbance in
fitting the fluxes of the lines reaches a threshold (here equal to that of
sinc$^2$ profiles at the Rayleigh criterion). The second criterion assumes
that two spectrographs have equal resolving powers if the wavelength error
in fitting a narrow spectral line is the same in each case (given equal
signal flux and noise power). The two criteria give similar results, and
give rise to scaling factors which can be applied to bring resolving power
calculated using the FWHM on to a consistent scale. The differences among
commonly encountered Line Spread Functions are substantial, with a
Lorentzian profile (as produced by an imaging Fabry-Perot interferometer)
being a factor of two worse than the boxy profile from a projected circle
(as produced by integration across the spatial dimension of a multi-mode
fibre) when both have the same FWHM. The projected circle has a larger FWHM
in comparison with its true resolution, so using FWHM to characterise the
resolution of a spectrograph which is fed by multi-mode fibres
significantly underestimates its true resolving power if it has small
aberrations and a well-sampled profile.
\end{abstract}
\maketitle %
\section{INTRODUCTION}
\label{sec:intro}
The spectral resolving power $R = \lambda/\delta\lambda$ is perhaps the most
important single property of a spectrograph. The wavelength
increment $\delta\lambda$ is the minimum separation for two spectral lines
to be considered as just resolved. The problem is that the definition of
$\delta\lambda$ is arbitrary, and inconsistent between various
usages. Classically the Rayleigh criterion was used, while in recent years
by far the most common practice has been to use the Full-Width at Half Maximum,
i.e. $\delta\lambda = {\rm FWHM}$. 

It is clear that there can be no fundamental definition of the minimum
resolvable wavelength difference $\delta\lambda$, because with arbitrarily
high signal/noise ratio, sufficiently fine sampling and a perfectly known
instrumental response function (here abbreviated as the Line Spread
Function, LSF\footnote{This departs from the usual interpretation of `Line
Spread Function' as the response of an optical system to a line source of
infinitesimal width. What is meant here is the system response to a
monochromatic input. It would be more accurately termed the `Spectral Line
Spread Function' (Spronck et al. 2013).}) an observed spectrum could be
deconvolved to any desired spectral resolution. What spectroscopists
understand by the `resolution' of an instrument is the smallest
$\delta\lambda$ which does not require (significant) deconvolution to
obtain spectral line strengths and locations (wavelengths). Lines of this
separation can be distinguished at moderate signal/noise levels. This
arbitrariness in the definition of $\delta\lambda$ has always been
recognised, from the early use of the Rayleigh criterion.

There is in principle no problem with an arbitrary definition of
$\delta\lambda$ and hence $R$, provided it is consistent between various
systems that are to be compared. Thus meaningful comparisons could be made
using $\delta\lambda = {\rm FWHM}$ {\it provided} that the LSF has the same
form in each case. But the problem arises because this is not true: a
diffraction-limited slit spectrograph gives a ${\rm sinc}^2$ profile, a
projected multi-mode circular fibre feed gives a boxy profile (a half
ellipse), a Fabry-Perot etalon with high finesse gives a Lorentzian
profile, a single-mode fibre or waveguide will give a Gaussian profile, and
a LSF with significant aberrations may resemble a Gaussian but in general
will have its own unique form.  It is when comparing resolving power
between instruments with different forms of LSF that inconsistency arises,
and as shown below the inconsistency can exceed a factor of two in
resolving power.  This is a significant error in the context of scientific
requirements for resolution, e.g. in stellar abundance studies. Moreover,
resolving power is typically one of the formal specifications of a
spectrograph, yet without a description of the LSF and the way
$\delta\lambda$ is to be measured, any requirement on $R$ is necessarily
imprecise in its meaning. Likewise, the concept of signal/noise per
resolution element is vague because the `resolution element' is not well
defined.

Inconsistencies also occur between the well-known formulas for theoretical
resolving power: 

a) $R = mN$ for a diffraction-limited slit spectrograph with uniform
illumination of all grating lines ($m$ = diffraction order, $N$ = number of
illuminated lines) assumes a ${\rm sinc}^2$ LSF and the Rayleigh criterion,
i.e. the maximum of a spectral line of wavelength $\lambda$ occurs at
the same position on the detector as the first zero of the line at $\lambda
+ \delta\lambda$.

b) $R=mF$ for a Fabry-Perot instrument ($m$ = order of interference, $F$ =
etalon finesse) assumes separation of the two Lorentzian LSFs by their
FWHM.

c) $R=2 b \tan\theta_i / (D \tan \theta_s)$ for a slit-limited spectrograph
used in Littrow configuration ($b$ = collimated beam diameter, $\theta_i$ =
grating incidence angle, $D$ = telescope diameter, $\theta_s$ = slit width
in angular measure on the sky) assumes rectangular LSFs (i.e. perfect
images of a uniformly illuminated slit) and two lines are regarded as just
resolved when the two slit images just touch.

There is thus a need to provide a more consistent definition of resolving
power, so that comparisons can be made with better precision.

In this paper I first illustrate the problem by comparing various LSF forms
with two lines separated according to the various criteria that have been
proposed.  I then attempt to provide a consistent definition of resolution
across different LSF forms. 

The influence of sampling of spectra into discrete pixels is important in
practice, but will be considered separately in a later work. For the
present paper, sampling issues will be avoided by using a
sufficiently large number of pixels so that profiles are effectively
continuous. This will keep the discussion focused on the issue of
resolution itself. The discussion here will be confined to 1-dimensional
spectra, e.g. after processing to integrate over the spatial direction of a
raw 2-dimensional data set.
\begin{figure*}
\begin{center}
\includegraphics*[angle=90, scale=0.47]{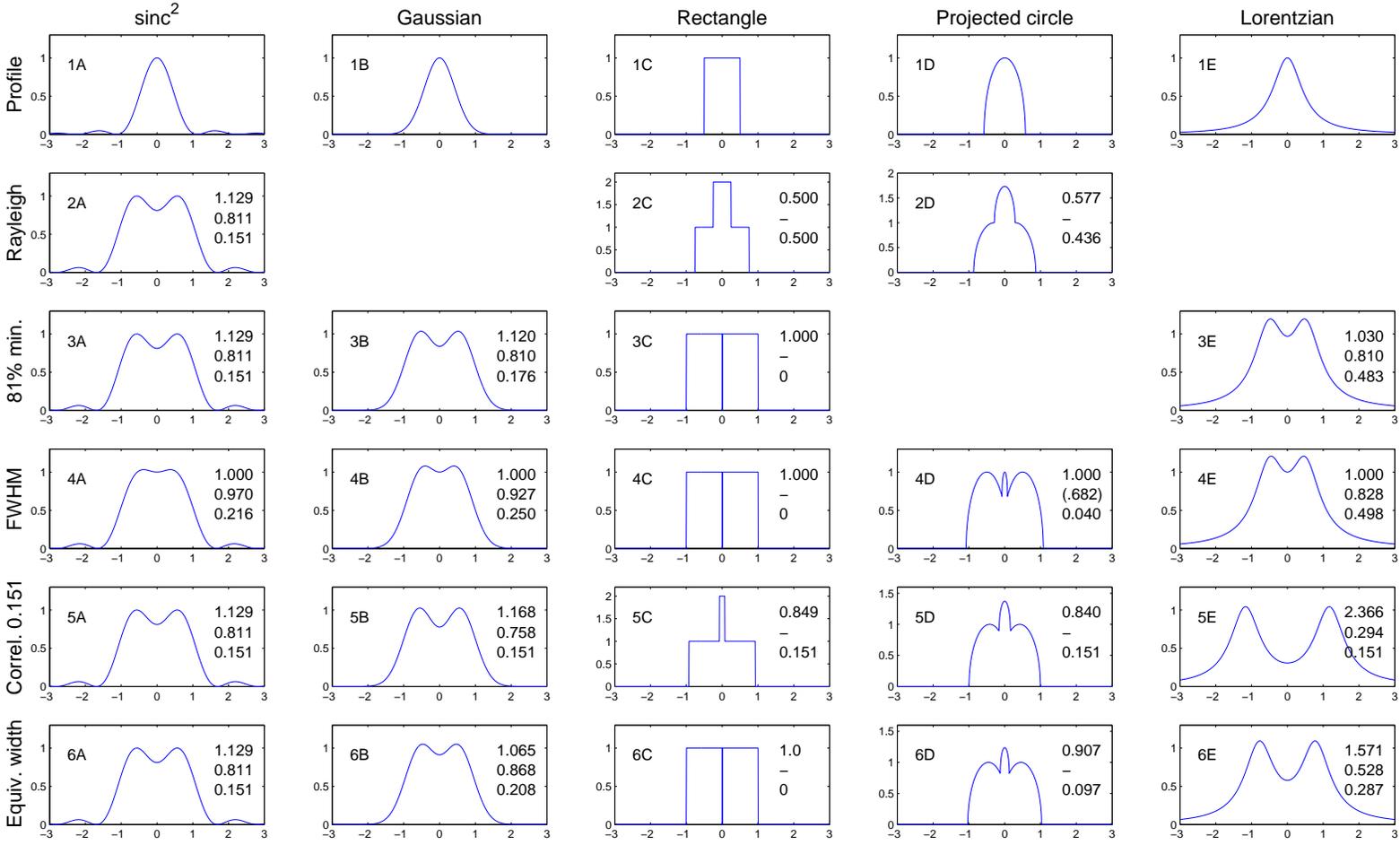}
\caption{Five Line Spread Function forms (top row, each shown with unity
  FWHM) with five different resolution criteria illustrated in the rows
  below. The three numbers at the right hand side of each panel are (from
  the top) the separation of the two peaks as a multiple of the FWHM, the
  relative minimum between the two peaks, and the autocorrelation at the
  separation shown (normalised to a peak of 1.0).}\label{fig:bigplot}
\end{center}
\end{figure*}
\section{RESOLUTION CRITERIA COMPARED} 
\label{sec:criteria}
Figure \ref{fig:bigplot} compares the different LSF profiles used in this
work and the various resolution criteria.  There are a number of points to
note from this Figure. Taking the rows in order:

1) The top row shows a single spectral emission line of each LSF form. The
   ${\rm sinc}^2$, rectangular and Lorentizian LSFs were introduced
   above. The Gaussian is often used as a general form of smooth profile,
   perhaps caused by many small errors and aberrations smoothing the ideal
   profile and combining via the Central Limit Theorem to give a Gaussian
   distribution. The projected circle profile in column D applies to the
   case of a multimode fibre, which presents a uniformly illuminated
   circular image at the spectrograph entrance. When integrated over the
   spatial direction and presented as a profile along the wavelength axis,
   it has the form of a half-ellipse. (This is an Abel transform; see
   e.g. Bracewell 1995 p 367.)

2) This and the subsequent rows show a pair of identical lines separated
   according to various criteria. The three numbers towards the right hand
   side of each panel show the separation/FWHM, the local minimum and the
   value of the autocorrelation at the separation shown. Panel 2A shows the
   classical Rayleigh criterion separation of two ${\rm sinc}^2$
   profiles. The local minimum between the peaks is 81.1\% of the peak
   height. To many spectroscopists this does indeed represent what is meant
   by two lines being just resolved. But the separation is 1.129 $\times$
   FWHM, illustrating the inconsistency of the two criteria. The Rayleigh
   criterion, where one peak is placed over the zero of the other profile,
   cannot be used for the Gaussian or Lorentzian profiles which do not have
   a zero. For the projected circle the boxy profile, with slope increasing
   as the edge of the profile is approached, produces the central spike in
   the sum as seen in all of panels 2D to 6D. In practice aberrations and
   pixelisation will remove this to some extent, but its effects must still
   be considered.

3) The Rayleigh criterion can be generalised by taking its local minimum
   of 81.1\% as the defining criterion. This can be applied to all except
   the projected circle, due to its central spike.

4) The FWHM is the most-used criterion nowadays. But as panels 4A and 4B
   show, for the ${\rm sinc}^2$ and Gaussian profiles the resulting blended
   profile is not well resolved. For the ${\rm sinc}^2$ profile (4A) the
   local minimum is 97\% of the peak, which does not accord with the common
   understanding of resolution.  A Gaussian profile (4B) is only a little
   better.  The projected circle (4D) has an overall flux deficit between
   the peaks but a central spike at the midpoint. In practice the result
   will depend on the degree of smoothing and pixelisation. For the
   Lorentzian profile (4E) the relative minimum is well seen but only with
   good signal/noise, due to the substantial overlap of the line wings
   (note the high autocorrelation of 0.498).

5) Again using the ${\rm sinc}^2$ profiles separated at the Rayleigh
   criterion as a standard, this row takes the resulting autocorrelation
   value of 0.151 and uses it as a criterion for two lines to be just
   resolved. Due to the high wings of the Lorentzian, it requires a
   separation of 2.366 $\times$ FWHM to meet this criterion (panel 5E).

6) The equivalent width (area/height) has been proposed to meet some of the
   above objections (e.g. Jones et al. 1995, 2002). For the ${\rm sinc}^2$
   profiles, a separation of 1.0 equivalent width is extremely close to the
   Rayleigh criterion. For other profiles it also gives reasonable results.

The conclusion from Figure \ref{fig:bigplot} is that none of the separation
criteria shown is clearly superior for all LSF forms, and in particular
the FWHM is a poor indicator of resolution for the important cases of
smooth sinc$^2$ or Gaussian profiles.  For reference, the main properties
of the LSF functional forms discussed in this paper are given in Table
\ref{tab:props}.
\begin{table*}[h]
\begin{center}
\caption{Line Spread Function properties$^a$}\label{tab:props}
\begin{tabular}{lccccc}
\hline  & formula & Peak & FWHM & EW & Z \\
\hline \smallskip \\
${\rm sinc}^2$ & $ \frac{0.8859}{\Gamma} (\frac{\sin \pi
  \frac{0.8859}{\Gamma}x}{ \pi \frac{0.8859}{\Gamma}x})^2 $ &
$ \frac{0.8859}{\Gamma}$ & $\Gamma$ & $\frac{\Gamma}{0.8859}$ & 
 $0.4289 \Gamma$ \medskip\\
Gaussian & $\frac{1}{\sigma \sqrt{2\pi}} \exp(-\frac{x^2}{2\sigma^2})$ &
$\frac{1}{\sigma \sqrt{2\pi}}$ & $\sigma. 2\sqrt{2 \ln 2}$ & $ \sigma
\sqrt{2\pi} $ & $2\sigma / \sqrt{\pi}$ \medskip \\
Projected circle & $\frac{2}{\pi a^2} \sqrt{a^2 - x^2}$  [$|x| \le a$] &
$\frac{2}{\pi a}$ & $a \sqrt 3$ & $\frac{\pi a}{2} $ & - \medskip\\
Lorentzian & $\frac{1}{\pi} \frac{\Gamma /2}{(x^2 + (\Gamma/2)^2} $ &
  $\frac{2}{\pi \Gamma} $ & $\Gamma$ & $\frac{\pi \Gamma}{2} $ 
 & $0.6367 \Gamma$ \medskip\\
\hline
\end{tabular}
\medskip\\
$^a$All formulas are normalised to unit area under the profile.\\
\end{center}
\end{table*}

\section{A CONSISTENT RESOLUTION CRITERION}
\label{sec:consistent}
This paper aims to present criteria by which resolving power can be more
meaningfully compared across LSFs of different functional forms. Two
approaches have been taken. In section \ref{sec:wavelength} a criterion
based on wavelength accuracy will be given. However the first criterion, to
be discussed in this section, is developed by recognising that what an
astronomer means by two close spectral lines being resolved is that the two
can be seen separately and can have their strengths and positions
(wavelengths) measured without undue influence of one on the other. There
will still be an arbitrary definition of what constitutes `undue'
influence, but the aim is to ensure that there is only {\it one} arbitrary
definition and that all other measures are consistent with it. The
influence of one spectral line on another is measured by its effect in
increasing the noise in measurement of the flux of the line.

The procedure to use this method was to generate LSFs of various functional
forms, with two equal strength peaks at separations varying from 0.8 to
2.0 $\times$ FWHM, add noise to them and then perform least squares fits to
extract the positions and strengths of the two peaks\footnote{All
  computations were performed using MATLAB (www.mathworks.com.au)}. Importantly, the
width of each peak was treated as known rather than as a further variable to
fit. This was done for two reasons: (1) at the ultimate closest resolvable
approach of two spectral lines it is recognised that the issue is to
separate two unresolved lines. It is well known that if the line width is itself
resolved then lines would have to be further apart to be properly
resolved. This is not what `spectral resolving power' is taken to mean. (2)
Once two lines begin to blend, in the presence of noise the fitting process
would be likely to result in one broadened line rather than two partly
blended lines.

\begin{figure}[h]
\begin{center}
\includegraphics[scale=0.47, angle=0]{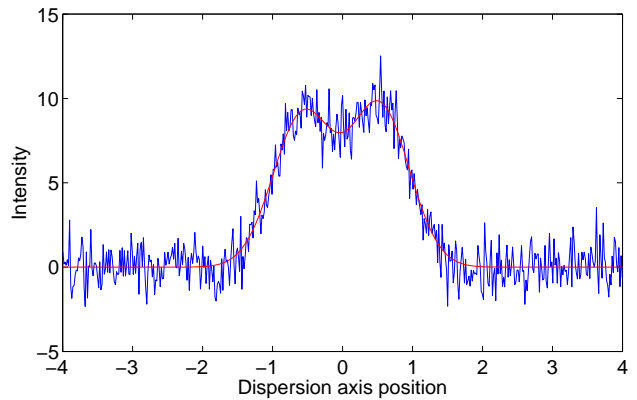}
\caption{Example of dual peak fitting. Blue line: two Gaussians each with
  peak 9.3941 (area 10.0), unity FWHM, at a separation of $1.1 \times $
  FWHM, with 62.6 samples over a FWHM, and subject to independent Gaussian
  distributed noise with standard deviation 1.0 in each sample. Red curve:
  the least squares fit to two Gaussians. This plot shows one of 4000
  realisations at one of 25 peak spacings.}\label{fig:example_dual_peak}
\end{center}
\end{figure}

\begin{figure}[h]
\begin{center}
\includegraphics[scale=0.455, angle=0]{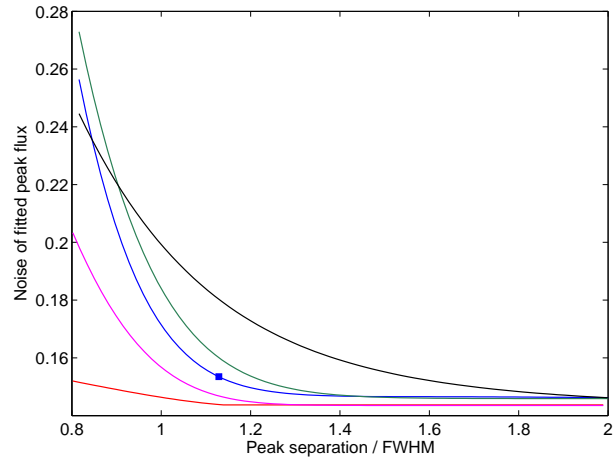}
\caption{The variation of $\sigma_{\rm flux}$ vs separation of two peaks,
  for five different LSF forms. From highest to lowest at peak separation =
  1.0 the curves are: black - Lorentzian; green - Gaussian; blue -
  sinc$^2$; magenta - projected circle convolved with a Gaussian (see
  Section \ref{sec:proj_circ}); red - projected
  circle. The blue square on the sinc$^2$ curve indicates the Rayleigh
  criterion separation.}\label{fig:sig_flux_sepn}
\end{center}
\end{figure}

Figure \ref{fig:example_dual_peak} shows an example of two lines, with
added noise, and the least squares fits. The simulations were performed
using the same noise power within the FWHM for each LSF form. This is an
unavoidably arbitrary choice of noise power normalisation, but it does not
influence the results to be derived from these simulations.  The different
LSFs were normalised to the same total area i.e. flux (not peak). This
reflects the fact that total signal power in the spectral line is the
quantity of importance to astronomers.
Figure \ref{fig:sig_flux_sepn} shows the results of this process. For each
of the five LSFs shown (${\rm sinc}^2$, Gaussian, Lorentzian, projected
circle and projected circle convolved with a Gaussian), a large number of
trials (4000) was done at each of 25 separations from 0.8 to 2.0 $\times$
the FWHM. From each set of 4000 trials the standard deviation of the least
squares fitted flux was found. The smooth curves shown are semi-empirical
model fits to the data of standard deviation versus peak separation and are
used to smooth out irregularities due to random fluctuations. The
functional form fitted was:
\begin{equation}
\sigma_{\rm fitted\ flux} = C \ ({\rm autocorrelation}(B^2(x)))^\gamma +
\sigma_{\rm flux,iso}.
\end{equation}
where $B(x)$ is the LSF function, $x$ being the independent variable along
the dispersion axis. Two free parameters, $C$ and $\gamma$, were adjusted
to fit the simulation results for each LSF and in all cases gave a very
good fit, within the residual fluctuations. The values of $\sigma_{\rm
flux,iso}$ were obtained using equation \ref{eqn:sig_flux} below.

At large separations the standard deviations approach the value obtained
for an isolated peak, i.e. by this criterion the lines are not influencing
each other, and are fully resolved.  However the Lorentz profile has such
broad wings that it has not yet reached a constant level at the separation
of 2.0 FWHM. The Lorentz profile shows the effect of one peak disturbing
another (i.e. increasing its noise) at substantially larger separations
than the other LSFs, when measured in multiples of the FWHM.

The ${\rm sinc}^2$ and Gaussian LSFs show very similar curves in Figure
\ref{fig:sig_flux_sepn}, consistent with the fact that both are peaked
functions which drop smoothly and rapidly towards zero.

The projected circle LSF has a very different curve of $\sigma_{\rm flux}$
vs separation. There is no effect at all of one peak on the other until
they begin to touch, at $2/\sqrt 3 \times$ FWHM = 1.1547 $\times$ FWHM. At
smaller separations there is some interaction but it is very small because
the profiles are convex with such steep sides. Figure
\ref{fig:sig_flux_sepn} also includes a curve for a projected circle LSF
convolved with a Gaussian of width such that the final FWHM is a minimum
(see section \ref{sec:proj_circ}).

Figure \ref{fig:sig_flux_sepn} makes clear that different LSF functional
forms do indeed have very different properties as regards the mutual
effects of two lines, and to simply use the FWHM as a resolution criterion
is a poor indicator of spectral resolution as it affects line finding and
fitting. It is also clear that the Lorentzian profile will give poor
resolution at a given separation in FWHMs, while the projected circle is
exceptionally good.

The data in Figure \ref{fig:sig_flux_sepn} can be used to derive scaling
factors to quantitatively compare different LSFs. The method used here was
to take a ${\rm sinc}^2$ profile separated according to the Rayleigh
criterion as the standard of `just resolved' spectral lines. This leads to
a  $\sigma_{\rm flux}$ value increased by a factor of 1.0514 compared with
its limiting value at large separations (i.e. for isolated peaks). Other
LSF forms will thus be considered to be just resolved when their
$\sigma_{\rm flux}$ values are likewise increased by $1.0514 \times$ over
the value at large separations. Defining a resolving power according to
this criterion:
\begin{equation}
 R_{\sigma{\rm flux}} = R_{\sigma_{\rm flux}=1.0514\times\sigma_{\rm flux,iso}} = 
 R_{\rm FWHM}/ \alpha
\end{equation}
where $\alpha$ is the separation/FWHM required to achieve the above
criterion, and $\sigma_{\rm flux,iso}$ is the standard deviation of a flux
measurement for an isolated peak (equation \ref{eqn:sig_flux}), the values
given in Table \ref{tab:scaling_factors} are obtained.

Although the ${\rm sinc}^2$ profile was used as the standard for
resolution, its value of $\alpha$ is not unity because the Rayleigh
criterion corresponds to a peak separation of 1.129 $\times$ FWHM.  The
$\alpha$ values show how much the resolving powers determined by the
present criterion of equal disturbance in peak fitting due to an adjacent
line differ from those based simply on the FWHM. As expected, the
Lorentzian is the worst, with an $R_{\sigma {\rm flux}}$ only 59\% of its
$R_{\rm FWHM}$ while the projected circle is the best, with $R_{\sigma{\rm
flux}}$ exceeding $R_{\rm FWHM}$ by 20\%. The convolved projected circle is
a more realistic case (to be discussed in Section \ref{sec:proj_circ}) and
its resolving power, while less than the exact projected circle, is still
substantially greater than a Gaussian or ${\rm sinc}^2$.
\begin{figure*}
\hspace*{-3cm}
\begin{center}
\includegraphics*[angle=90, scale=0.47]{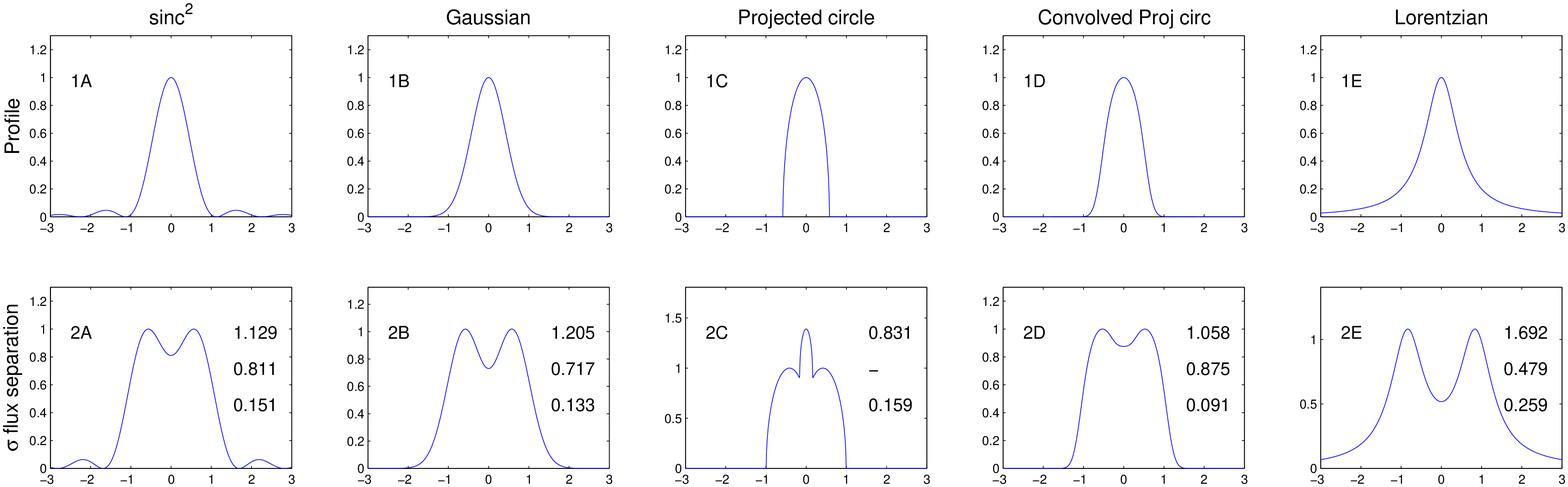}
\caption{Five LSF functional forms (top row). The convolved projected
  circle is the configuration with minimum final FWHM (FWHM$_{\rm
  projected\ circle}$ = 1.0532, FWHM$_{\rm Gaussian}$ = 0.3432, FWHM$_{\rm
  final}$ = 1.000).  The first row shows the single LSFs, while the second
  row shows a pair of peaks of the corresponding LSF separated according to
  the criterion $\sigma_{\rm flux}=1.0514\times\sigma_{\rm flux,iso}$
  introduced in Section \ref{sec:consistent}. The three numerical values at
  the right hand side of each panel are as for Figure
  \ref{fig:bigplot}.}\label{fig:medplot}
\end{center}
\end{figure*}
\begin{table}[h]
\begin{center}
\caption{Resolution element scaling factors}\label{tab:scaling_factors}
\begin{tabular}{lcc}
\hline LSF form & $\alpha$ & $\beta$ \\
\hline \\ [-0.2 cm]
${\rm sinc}^2$ & 1.129 & 1.129 \\
Gaussian & 1.21 & 1.127 \\
Lorentzian & 1.70 & 1.605 \\
Projected circle & 0.83 \\
Projected circle (conv) & 0.95 & 0.943 \\
\hline
\end{tabular}
\end{center}
\end{table}
Figure \ref{fig:medplot} shows profiles presented in the same style as
Figure \ref{fig:bigplot} but with row 2 showing various LSF types with two
peaks separated according to the criterion $\sigma_{\rm
  flux}=1.0514\times\sigma_{\rm flux,iso}$. These show the separations which are
regarded as `just resolved' according to the criterion introduced here.

\section{LINE PARAMETER UNCERTAINTIES}
\label{sec:uncerts}
Before introducing a second method of quantifying resolving power, it is
necessary to review the formulas for uncertainties in the flux and position
(wavelength) of a single spectral line peak.

When the width of the peak is known and only the amplitude (flux) and
position (wavelength) are fitted by least squares, and assuming a
symmetrical LSF form, Clarke et al (1969) give the formulas:
\begin{equation}
\sigma_{\rm flux,iso} = \sigma / \sqrt{\sum{B^2}} \label{eqn:sig_flux}
\end{equation}
\begin{equation}
\sigma_{\lambda,{\rm iso}} = \sigma / \Big( {\rm pk} \sqrt{\sum{(B')^2}}\Big)
\label{eqn:sig_x}
\end{equation}
In these formulas $\sigma$ is the rms noise in each wavelength channel and
is assumed to be the same for all channels. The summation is over all
wavelength channels contributing to the profile. The LSF function is $B$,
and $B'$ denotes its derivative with respect to wavelength. Note that in
these equations $B$ is normalised to a peak of 1.00, and the `pk' in eqn
\ref{eqn:sig_x} is the peak flux of the response whose $\sigma_\lambda$ is
to be found.  These formulas have been verified by Monte Carlo tests
and show that the precision in finding the strength of a peak depends most
on the values where the intensity is greatest, while the precision in
location of the peak depends on the regions of greatest slope.

It is not appropriate in this paper to consider a detailed noise model
where one would take into account shot noise from both the object and the
background sky, as well as read-out noise and dark noise. Instead, it will
suffice to use the above assumption of constant noise in all channels. The
results are thus most directly applicable to spectra that are background (or
read-out noise) limited but serve as a guide for other noise models as
well. They can also be applied to absorption lines, especially those that
do not depress the continuum by a large fraction.

In the present work a large number of channels (pixels) have been used,
e.g. 62.5 or 100 across the FWHM, to avoid the issue of sampling
effects. However, for the projected circle the gradient $B'$ becomes
infinite as the intensity drops to zero, and the sum in Equation
\ref{eqn:sig_x} would always be dominated by the edge pixels (see Section
\ref{sec:proj_circ}).  Hence this case is omitted here. The more realistic
convolved projected circle avoids this problem.

\section{A SECOND RESOLUTION CRITERION}
\label{sec:wavelength}
The second method to be considered originates from a somewhat independent
property of high resolving power, namely the ability to measure accurate
positions (wavelengths) of unresolved spectral lines. Thus two
spectrographs can be considered as having equal resolving power if they
give the same wavelength accuracy despite their different LSF forms,
assuming the noise power per wavelength interval remains constant and equal
total fluxes are received in both cases.

To compare resolving powers using this criterion, there is no need to
perform noise simulations as in section \ref{sec:consistent} but instead
equation \ref{eqn:sig_x} can be used as follows.

Define
\begin{equation}
Z = \frac{1}{\int_{-\infty}^{+\infty}(B')^2 d \lambda}.
\label{eqn:Z}
\end{equation}
$Z$ is a type of width measure of a LSF, which will be referred to as the
`noise width', given its role in calculating $\sigma_\lambda$. For
empirically determined LSFs $Z$ will generally be calculated numerically as
\begin{equation}
Z \simeq \frac{1}{\Delta \lambda \sum{(B')^2}}.
\label{eqn:Z_sum}
\end{equation}
where $\Delta \lambda$ is the channel width in the summation. Values of $Z$
for the LSF types discussed here are included in Table \ref{tab:props}.

Equation \ref{eqn:sig_x} can now be written as
\begin{equation}
\sigma_\lambda = \sigma Z^{\frac{1}{2}} \Delta \lambda^\frac{1}{2} / {\rm pk} 
\label{eqn:sig_x_2}
\end{equation}
where again $\sigma$ is the rms noise in the channel of width $\Delta
\lambda$ and the subscript `iso' has been omitted because all profiles considered
in this section are single.

The basis of this second resolution criterion is that $\sigma_{\lambda,{\rm
LSF}}$ of any LSF will be equated to $\sigma_{\lambda,{\rm sinc}^2}$, with the
condition that the two profiles have equal total fluxes (not equal peak
values). 

The condition for equal total fluxes is simply
\begin{equation}
{\rm pk}_{{\rm sinc}^2} = {\rm pk}_{\rm LSF} \times \frac{{\rm EW}_{\rm
    LSF}}{{\rm EW}_{{\rm sinc}^2}}
\label{eqn:EW}
\end{equation}
where $EW$ stands for the equivalent width.
Equating the $\sigma_\lambda$'s for the given LSF and for sinc$^2$ and
using the values of $Z$ and $EW$ for sinc$^2$ from Table \ref{tab:props}
there follows
\begin{equation}
{\rm FWHM}_{{\rm sinc}^2} = 1.2231 Z_{\rm LSF}^{\frac{1}{3}} {\rm EW}_{\rm
  LSF}^{\frac{2}{3}}.
\label{eqn:FWHM}
\end{equation}
This is the FWHM of a sinc$^2$ profile which would have the same wavelength
noise error as the actual LSF being examined. If the value is large it
means that a wide sinc$^2$ could give accuracy equal to the LSF i.e. the
LSF is poor (e.g. a Lorentzian). If the FWHM$_{{\rm sinc}^2}$ is narrow it
means that a high resolution sinc$^2$ is needed to equal the accuracy of a
good LSF, e.g. the convolved projected circle.

The final step is to form the ratio of this calculated FWHM$_{{\rm sinc}^2}$
with that of the actual LSF and scale it by a factor 1.129 which will make
the final scaled resolving powers consistent with the Rayleigh
criterion for sinc$^2$ profiles. This gives
\begin{equation}
\beta = 1.3809 \ Z_{\rm LSF}^{\frac{1}{3}} {\rm EW}_{\rm LSF}^{\frac{2}{3}}/
      {\rm FWHM}_{\rm LSF}.
\label{eqn:beta}
\end{equation}

Values of $\beta$ for the standard LSF forms are included in Table
\ref{tab:scaling_factors}, except for the projected circle where the
infinite gradient limit makes the calculation invalid. Values are quite
similar to the $\alpha$ scaling factors derived in section
\ref{sec:consistent}.

The interpretation of $\beta$ is that 
\begin{equation}
\delta \lambda_{\sigma \lambda} = \beta \ {\rm FWHM}_{\rm LSF}
\label{eqn:del_lam}
\end{equation}
is the effective $\delta \lambda$ which should be used in place of the FWHM
in order to calculate resolution on a scale consistent with the Rayleigh
criterion for a sinc$^2$ profile. Thus
\begin{equation}
R_{\sigma \lambda} = \frac{1}{\beta} \frac{\lambda}{{\rm FWHM}_{\rm LSF}}
\label{eqn:R}
\end{equation}
is the resolving power on this consistent scale.

This criterion will be easier to use in practice than the $\sigma_{\rm
flux}$-based criterion of section \ref{sec:consistent}. For an empirically
determined LSF, for example resulting from ray tracing of a spectrograph
design, one would need to interpolate the LSF to a fine sampling interval,
and smooth out any fine structure artefacts from the LSF calculation
(e.g. from a finite number of traced rays), then use equation
\ref{eqn:Z_sum} to find the noise width and also find the FWHM and
equivalent width (area/peak). Then equation \ref{eqn:beta} can be used to
find the scaling factor which is finally applied in equation
\ref{eqn:R}. In the case of an asymmetric LSF the more general form of
equation \ref{eqn:sig_x} given by Clarke et al. (1969) eqn
(A7)\footnote{Note there is a typographical error in their eqn (A7), where
$\sum{(B_0B'_0)^2}$ should be replaced by $(\sum{B_0B'_0})^2$} should be
used, although the corrections for asymmetry are small.
\section{PROJECTED CIRCLE LSF}
\label{sec:proj_circ}
The projected circle LSF is important in practice and has very different
properties compared with other forms, and so warrants further discussion.
The use of multi-mode fibres to feed images to a pseudo-slit in a
spectrograph is increasingly common. Taking the fibre exit face as a
uniformly illuminated circle (a good approximation given the spatial
scrambling produced by transmission along the fibre), when its image has
been integrated over the spectrograph's spatial direction, the result will
be the projected circle as illustrated in panel 1D of Figure
\ref{fig:bigplot}. It 
differs markedly from the sinc$^2$, Gaussian and Lorentzian forms in that
the projected circle LSF approaches the x-axis with infinite slope. This
convex-outwards form results in the formation of a central spike when two
such LSFs overlap, as in Figure \ref{fig:bigplot}.

Interestingly, the projected circle line profile also results from Doppler
broadening of an intrinsically narrow line in a rapidly rotating star. This
is because the radial velocity is constant along strips parallel to the
rotation axis, and the flux at any one wavelength is due to an integration
along such a strip, i.e. a projection. The effects of the very steep sides
of such a profile have been noted, and Dravins (1992) drew attention to the
sharp spectral features which could appear at wavelengths where no spectral
line is present, i.e. the central spikes as seen in Figure
\ref{fig:bigplot}. He also noted that information about the true stellar
spectrum could be obtained regarding features considerably narrower than
the FWHM of the full broadened profile - this is again due to the steep
sides, which lead to the central spike being narrow and easily smoothed out
(in this case by intrinsic line width in a stellar spectrum).

As shown in Section \ref{sec:consistent} the lack of wings of the projected
circle LSF result in minimal noise interaction of two close lines, i.e. its
effective resolving power is substantially higher than its FWHM would
suggest.  

\begin{figure}[h]
\begin{center}
\includegraphics[scale=0.41, angle=0]{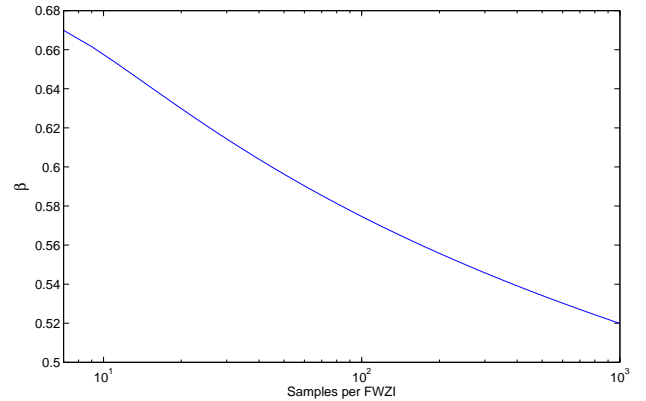}
\caption{Variation of $\beta$ for a projected circle LSF as a function of
  the number of samples across the Full Width to Zero Intensity
  (FWZI).}\label{fig:proj_circle_beta}
\end{center}
\end{figure}

The pure projected circle LSF cannot be directly compared with
other LSFs as regards wavelength uncertainties, because of the infinite
slopes. This means that however fine the sampling may be, the $\beta$ value
will still depend on the sampling interval. This is illustrated in Figure
\ref{fig:proj_circle_beta} which shows $\beta$ dropping approximately
logarithmically with increasing sampling frequency. The values of $\beta$
shown are all substantially less than any of those in Table
\ref{tab:scaling_factors}. Even with some blurring due to aberrations a
well-sampled LSF resembling the projected circle will have much higher
wavelength accuracy than a Gaussian-like peak of the same FWHM.

\begin{figure}[h]
\begin{center}
\includegraphics[scale=0.47, angle=0]{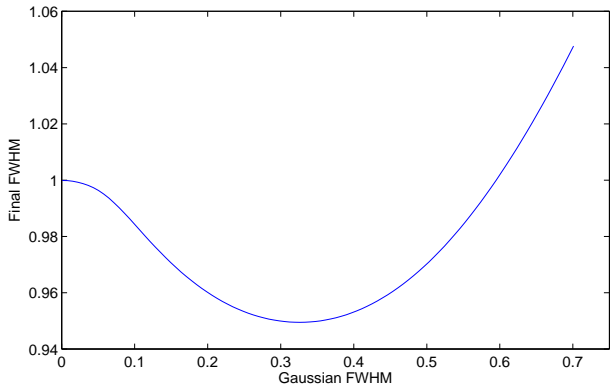}
\caption{Final FWHM after convolving a projected circle LSF of unity FWHM
  with a Gaussian of FWHM as given by the horizontal axis.}\label{fig:conv_width}
\end{center}
\end{figure}

\begin{figure}[h]
\begin{center}
\includegraphics[scale=0.47, angle=0]{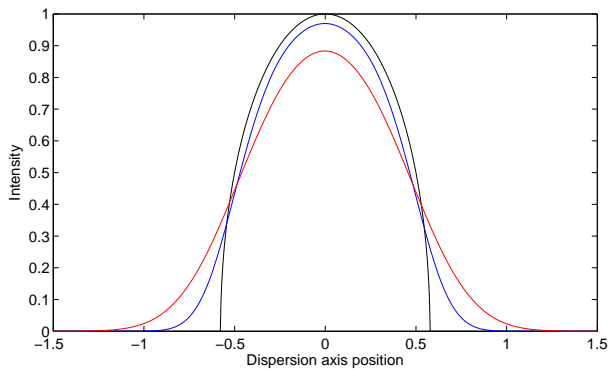}
\caption{Three of the resulting curves from the convolutions of Figure
  \ref{fig:conv_width}. The curves from highest to lowest at the peak are:
  black: pure unconvolved projected circle; blue: Gaussian FWHM = 0.3259
  gives the minimum final FWHM of 0.9494; red: Gaussian FWHM = 0.595
  results in a final FWHM of 1.00.} \label{fig:three_conv}
\end{center}
\end{figure}

One of the peculiarities produced by the convex boxy shape of this LSF is
that the FWHM is {\it reduced} by convolution with a Gaussian of moderate
width. This effect was noted in the design of the AAOmega spectrograph
(Saunders et al. 2005). This is another illustration of the inadequacy of
FWHM as a measure of resolution, since one would not claim that convolution
of the LSF by spectrograph aberrations increases the resolving power.
Figure \ref{fig:conv_width} illustrates this behaviour, using a projected
circle LSF of FWHM = 1.00 convolved with Gaussians of various FWHMs up to
0.7. The resulting FWHM drops by as much as 5\%, when the Gaussian FWHM =
0.3259, before rising again as the Gaussian convolving function is further
broadened. Figure \ref{fig:three_conv} shows three of the profiles: the
pure projected circle; the case of the minimum final FWHM, and the case of
Gaussian FWHM = 0.595 which restores the final FWHM to 1.00, albeit with a
very different LSF form compared with the initial projected circle. The
case of the minimum final FWHM was used as the example of a convolved
projected circle in Table \ref{tab:scaling_factors} and Figures
\ref{fig:sig_flux_sepn} and \ref{fig:medplot}.
\section{CONCLUSIONS}
\label{sec:conclusions}
The analysis above has shown that characterising the resolution $\delta
\lambda$ of a spectrograph by its instrumental FWHM is a poor measure
because it fails to take fully into account the variation among different Line
Spread Function forms of the quantities which matter most in spectroscopy
- namely the disturbance which a spectral line causes to a near neighbour,
or the accuracy with which a single line's wavelength can be
measured. Using these two criteria, a very different picture emerges, as
shown by the $\alpha$ and $\beta$ scaling factors in Table
\ref{tab:scaling_factors}. There is more than a factor of two difference in
resolving power between the best and worst LSFs (with identical FWHM) when
resolving power is measured on a consistent scale. 

Comparing the various resolution criteria shown in Figure \ref{fig:bigplot}
with the $\sigma_{\rm flux}$-based criterion of Figure \ref{fig:medplot}
shows that the Equivalent Width is the one that comes closest to matching
the consistent resolution scale introduced here. But the match is not
exact, with a significant difference in the case of the Gaussian LSF.

The Lorentzian LSF's broad wings greatly increase its effective $\delta
\lambda$ and hence reduce the resolving power of an instrument with this
LSF well below the value given by the FWHM. It is well known by users of
imaging Fabry-Perot instruments, for example, that this LSF makes the
instrument unsuitable for absorption line studies, because a line core is
influenced by convolution with continuum fluctuations over a substantial
wavelength range. Here, this influence has been quantified and the
Lorentzian's low relative resolving power explicitly demonstrated.

Conversely, the projected circle, even after smoothing by significant
aberrations, has a steep-sided form which gives substantially higher
resolving power than its FWHM would suggest. Gaussian and sinc$^2$ profiles
have properties intermediate between these two extremes. But even they have
ambiguities at the 10-15\% level, with a pair of Gaussian profiles
requiring a separation of 1.129 $\times$ FWHM to achieve the 81\% relative
minimum of a generalised Rayleigh criterion. Either of the two
resolution element scaling factors can serve as a quality indicator for any
given LSF profile.

It is notable that the $\alpha$ and $\beta$ scaling factors in Table
\ref{tab:scaling_factors} are quite similar for a given LSF type, despite
the former being based on the additional error in fitting the flux of a
line caused by a near neighbour, while the latter is based on accuracy of
wavelength determination for isolated lines. This agreement strengthens the
case for using one of these resulting scaling factors to bring resolving
power of any spectrograph on to a consistent scale. In principle, the
`$\alpha$' factor, based on mutual disturbance in fitting a line is the
more appropriate in low to moderate signal/noise spectra, while the
`$\beta$' factor, based on wavelength accuracy, is the more appropriate for
high-resolution, high signal/noise work. But given the similarity of the
two factors and that the $\beta$ factor is much easier to calculate for a
general empirically-determined instrumental profile, the $\beta$ factor is
recommended as a suitable standard measure for comparison of resolving
power between different spectrographs. \\[2mm]

%
%
%
\noindent
{\bf REFERENCES}\\

\noindent
Bracewell, R.N. {\it Two Dimensional Imaging,} Prentice Hall 1995.\\[-1mm]

\noindent
Clarke, T.W., Frater, R.H., Large, M.I., Munro, R.E.B. and
  Murdoch, H.S. Aust. J. Phys. Astrophys. Suppl. 10, 3, 1969.\\[-1mm]

\noindent
Dravins, D. {\it High Resolution Spectroscopy with the VLT,}
  ed. Ulrich, M.-H., ESO Workshop No. 40, 55, 1992.\\[-1mm]

\noindent
Jones, A.W., Bland-Hawthorn, J. and Shopbell, P.L. ASP
  Conf. Ser. 77, 503, 1995.\\[-1mm]

\noindent
Jones, D.H., Shopbell, P.L. and Bland-Hawthorn, J. MNRAS
   329, 759, 2002.\\[-1mm]

\noindent
Saunders, W. AAO Newsletter No. 108, 8, August 2005. \\[-1mm]

\noindent
Spronck, J.F.P., Fischer, D.A., Kaplan, Z.A.,
  Schwab, C. and Szymkowiak, A. arXiv 1303.5792 2013.
%
%

\end{document}